\documentclass[ twocolumn,prd,amssymb,preprintnumbers,superscriptaddress,nofootinbib]{revtex4-1}
\usepackage{graphicx}
\usepackage{amsmath}
\usepackage{amsfonts}
\usepackage{mathtools}
\usepackage{color}
\usepackage{amssymb}
\usepackage{dcolumn}
\usepackage{bm}
\usepackage{braket}
\usepackage{microtype} 
\usepackage[linktoc=all]{hyperref}
\usepackage{cleveref}

\numberwithin{equation}{section}
\newcommand{\lag}{\mathcal{L}}
\newcommand{\ud}{\mathrm{d}}

\newcommand{\bel}[1] {\begin{equation}\label{#1}}
	\newcommand{\beal}[1] {\begin{eqnarray}\label{#1}}
		
		\newcommand{\be}{\begin{equation}}
			\newcommand{\bea}{\begin{eqnarray}}
				\newcommand{\ee}{\end{equation}}
			\newcommand{\eea}{\end{eqnarray}}
		
		\makeatletter
		\DeclareRobustCommand{\rcite}[1]{%
			\rcite@aux#1,\@nil{#1}%
		}
		\def\rcite@aux#1,#2\@nil#3{%
			\if\relax#2\relax
			Ref.~\cite{#3}%
			\else
			Refs.~\cite{#3}%
			\fi
		}
		\makeatother
		
\begin{document}
			\title{Bounds on EFT's in an expanding Universe}
			\author{Mariana Carrillo Gonz\'alez} 
             \email{m.carrillo-gonzalez@imperial.ac.uk}
			\email{}
			\affiliation{Theoretical Physics, Blackett Laboratory, Imperial College, London, SW7 2AZ, U.K }
			
			\date{\today}

   \preprint{Imperial/TP/2023/MC/03}

   \begin{abstract}
   We find bounds on the Wilson coefficients of effective field theories (EFTs) living in a Universe undergoing expansion by requiring that its modes do not propagate further than a minimally coupled photon by a resolvable amount. To do so, we compute the spatial shift suffered by the EFT modes at a fixed time slice within the WKB approximation and the regime of validity of the EFT. We analyze the bounds arising on shift-symmetric scalars and curved space generalizations of Galileons. 
   \end{abstract}
			
\maketitle		

\section{Introduction}
While the standard cosmological model has been extremely successful at explaining current observations, many mysteries remain. Some examples of such mysteries are the fundamental origin of the dark sector as well as some cosmological tensions \cite{Abdalla:2022yfr}, with the most prominent being the tension between the early and late time measurements of the acceleration of the Universe \cite{Riess:2020fzl,Planck:2018vyg}. It is expected that the knowledge of the fundamental theory of everything will resolve all of these unknowns, but in the absence of this, we can work with effective field theories (EFTs) that describe physics at low energies. The unknown description of the theory at high energies is encoded in the Wilson coefficients of these EFTs. From a bottom-up perspective, these Wilson coefficients have no a priori values, but it is well-known that certain values of these coefficients can lead to unphysical properties. 

In flat space, it is possible to define an S-matrix whose analytic properties are well known. By imposing physical principles, such as Lorentz invariance, analyticity, unitarity, and locality at all scales; one can write dispersion relations that allow us to bound the Wilson coefficients of a theory \cite{PhysRev.182.1400,Pennington:1994kc, Pham:1985cr,Adams:2006sv,deRham:2022hpx}. Applying this program to curved spacetimes and more specifically to an expanding Universe is not straightforward due to the lack of a globally defined S-matrix. This has been the subject of many recent explorations \cite{Baumann:2022jpr,Baumann:2015nta,Melville:2019wyy,deRham:2021fpu,Traykova:2021hbr,Kim:2019wjo,Herrero-Valea:2019hde,Ye:2019oxx,Grall:2021xxm,PhysRevD.77.023505,Green:2023ids,Albrychiewicz:2020ruh}. Nevertheless, these explorations are still in early stages and not nearly as developed as the flat space case. Here we analyze a novel method to obtain bounds on Wilson coefficients for EFTs living in a curved background where a subset of the Lorentz symmetries are broken. While the explicit examples we show correspond to fully covariant EFTs which undergo a spontaneous symmetry breaking due to the time-dependent scalar background, these techniques can be applied to EFTs defined directly on expanding backgrounds where time and spatial derivatives are treated separately. 

To constrain the values of the Wilson coefficients we will follow the techniques developed in \cite{CarrilloGonzalez:2022fwg,CarrilloGonzalez:2023cbf} for imposing bounds on EFTs in flat space based on causality requirements. The physical requirement that leads to these bounds is the causal propagation of modes around non-trivial backgrounds. For related studies on the effects of causality see \cite{deRham:2020zyh,Chen:2023rar,deRham:2021bll,Chen:2021bvg,Serra:2022pzl,Serra:2023nrn}. Explorations of causality on expanding spacetimes can be found in \cite{Bittermann:2022hhy,Dubovsky:2007ac,Baumgart:2020oby}. In Section \ref{sec:causal}, we review this setup for a spherically symmetric background. Then, Section \ref{sec:tdep} focuses on homogeneous backgrounds and the specific example of de Sitter. We apply these techniques to a shift symmetric scalar in Section \ref{sec:scalar} and to the de Sitter Galileon in \ref{sec:dSgal}. We compute bounds for operators up to mass dimension 12 with $4$ and $6$ fields. We shortly address the causal properties of potential terms in Section \ref{sec:pot} and conclude by discussing the results and future applications in Section \ref{sec:disc}. We also show explicit calculations at higher orders in the expansion in Appendix \ref{ap:ho} and for the de Sitter Galileon in Appendix \ref{ap:dSGal}. Last, we analyze physical requirements on the background external source in Appendix \ref{ap:source}.

\section{Causality around non-trivial backgrounds} \label{sec:causal}
Consider a generic effective field theory with a propagating mode $\phi$ satisfying the equation of motion
\begin{equation}
	\left(-\partial_t^2+\nabla^2+\sum_i \frac{g_i}{M^{p_i}}\mathcal{O}(\phi,\partial\phi,\dots,\partial,\dots)\right)\phi= g_{m} J \ ,
\end{equation}
where $J$ is an external source, $g_i$ the Wilson coefficients, $g_m$ the strength of the matter coupling, $M$ the cutoff of the theory, and $p_i \in \mathbb{Z}$ a power that gives each term the correct mass dimensions. We want to understand if the propagation around a non-trivial, localized background $\bar{\phi}=\bar{\phi}(x^\mu)$ sourced by $J$ is causal. To do so we consider a linearized perturbation $\varphi=\phi-\bar{\phi}$ around this background such that
\begin{equation}
	\left(-\partial_t^2+\nabla^2+\sum_i \frac{g_i}{M^{p_i}}\mathcal{O}(\bar{\phi},\partial\bar{\phi},\dots,\partial,\dots)\right)\varphi=0 \ . \label{eq:perteom}
\end{equation}	
This equation can be solved perturbatively using the WKB approximation assuming that the scale at which the perturbation varies is much smaller than that of the background. After interacting with the non-trivial background, the perturbation will experience a phase shift with respect to a free particle. This phase shift encodes the support of the retarded Green's function. To determine whether we have causal propagation, we will require that any incoming wave arrives at a given point before the outgoing wave leaves that point. To do so, let us consider a perturbation given by a wave packet traveling in a non-trivial background such that the outgoing wave reads
\begin{equation}
\varphi^\text{out}=\int \ud^3k A(k) \frac{1}{r} e^{i k r}e^{-i\omega t}e^{i 2 \delta} \ .
\end{equation}
The center of this wave packet is found by solving $\partial_k(k r-\omega t+2\delta)=0$. Thus, at a given fixed distance $r$ the interacting wave packet arrives at a time $t+2\partial_\omega\delta$ while a free one arrives at $t$, that is, it experiences a time delay given by 
\begin{equation}
\Delta T= 2\partial_\omega\delta \ .
\end{equation}
Alternatively, one can see this as a spatial shift of the center of the interacting wave packet that reads
\begin{equation}
    \Delta r=-2\partial_k\delta \ . \label{eq:Dr}
\end{equation}

We will assume that our spacetime has a chronology determined by a minimally coupled photon or massless high-energy mode. This means that, if our setup gives rise to a resolvable time/spatial advance, then we can construct closed timelike curves as in \cite{Adams:2006sv}. A different point of view has been assumed in \cite{Bruneton:2006gf,Bruneton:2007si,Babichev:2007dw} and allows for a large time advance without the consequence of closed time-like curves. The resolvability criteria is a consequence of the uncertainty principle which tells us that a time delay $|\Delta t |\sim \omega^{-1}$ cannot be accurately measured. Equivalently, a spatial shift $|\Delta r |\sim k^{-1}$ is not resolvable. This notion is encoded in Wigner's causal inequality:
\begin{equation}
k\Delta r<1 \ .  \label{eq:causal}
\end{equation}
For a recent review on this topic for see Section 2 of \cite{Mizera:2023tfe}. Here, we propose to use the exact non-relativistic version for our relativistic setup. The right-hand side (RHS) of this inequality is expected to be an order one number, but no strict derivation of this precise number, in analogy to the non-relativistic case \cite{Wigner:1955zz}, exits.  Hence, we will show how the bounds change under a change of this number. This test of causal propagation has been previously applied to leading order gravitational operators in FLRW backgrounds in \cite{deRham:2020zyh} without putting specific bounds on the Wilson coefficients. In this paper, we obtain bounds on higher-order Wilson coefficients of EFTs in an expanding Universe. Despite the EFT suppression on the higher-order operators, their contribution to the spatial shift can be enhanced by considering specific backgrounds that make the contribution of one operator larger but keep the contributions of all others of the expected EFT suppressed order. 

\section{Causality around homogeneous backgrounds} \label{sec:tdep}
We will now focus on the specific case of a homogeneous background $\bar{\phi}=\bar{\phi}(t)$. The perturbed field can be expanded as $\phi=f_k(t)e^{i{\mathbf{k}\cdot\mathbf{x}}}$ and the equation of motion, after field redefinitions that remove terms with $\dot{f}_k$,  is given by 
\begin{align}
	\ddot{f}_k(t)+W_k(t) f_k(t)=0 \ , \\
 \quad W_k(t)=\left(k^2 c_s^2(k,t)+V_\text{eff}  \right)\ ,
\end{align}
where $k=|\mathbf{k}|$, $c_s^2(k,t)$ is the effective speed of sound of the propagating mode and $V_\text{eff}$ its effective potential. We solve this equation with boundary conditions\footnote{The precise limit requires an $i \epsilon$ prescription to ensure convergence, that is, $t \rightarrow-\infty(1+i \epsilon)$.}
$\lim _{t \rightarrow-\infty} f_k(t)=\frac{1}{\sqrt{2 k}} e^{-i k t}$ using the WKB approximation. The leading order solution is given by
	\begin{equation}
		f=C \  e^{-i\int_{t_\text{in}}^t \sqrt{W_k(t)}\ud t}  \ ,
	\end{equation}
where $C$ is fixed by the boundary conditions. We can compute the phase shift by looking at the solution far away from the scatterer, that is, far away from the time-dependent localized background where the perturbation behaves as follows: $\lim\limits_{t\rightarrow\infty}f\propto\left(e^{2 i \delta_{\ell}} e^{-i k t}\right)$. For a background that varies over time scales $H^{-1}$ we have at leading order
\begin{equation}
	\delta\simeq - \frac{k}{2H} \int_{T_\text{in}}^{T_\text{out}}\left(c_s(k,T)-1\right)\mathrm{d} T \ .
\end{equation}
where we introduced the dimensionless time $T=H t$. The WKB expansion requires that the prefactor $k/H$ is large, while the validity of the EFT gives a small integrand; these two expansions compete to lead to a resolvable spatial shift.

\subsection{Fixed de Sitter backgrounds}
In this section, instead of working in flat spacetimes, we consider curved spacetimes, more specifically, de Sitter space. In de Sitter, the situation is similar to the one described above with the difference that at a fixed time slice a free scalar in de Sitter will experience a phase shift with respect to the plane wave in the far past due to the spacetime expansion. We will work in the Poincare patch of de Sitter which is described by the metric
\begin{equation}
    \ud s^2 = \frac{1}{H^2 \tau^2}\left(-\ud \tau^2+\ud {\bf x}^2\right) \ ,
\end{equation}
where $\tau<0$ is the conformal time. A scalar $\phi(T)=\Psi(T) f_k(T) e^{i{\mathbf{k}\cdot\mathbf{x}}} $ in this spacetime evolves following the equation of motion
\begin{equation}
	\ddot{f}_k(T)+\frac{k^2}{H^2}W_k(T) f_k(T)=0 \ , \label{eq:EOM} 
 \end{equation}
where the dimensionless time is now defined as $T=H \tau$, and $\Psi(T)$ is chosen such that it removes the friction term. We can obtain the phase shift using the WKB approximation at a constant time slice with conformal time $\tau_0<-\sqrt{2}/k$. We won't look at the phase shift at the boundary since at $\tau\geq-\sqrt{2}/k$ the WKB approximation breaks down. Thus, we always work at times before the mode crosses the Hubble horizon. 

To understand the effect of higher derivative operators and whether they lead to violations of causality, we take as a point of reference the support of the retarded Green's function of high energy modes propagating in de Sitter. We require that the retarded Green's function for the EFT does not have measurable support outside the region where the retarded Green's function for the high energy modes has support. In other words, we want to compute the phase shift experienced by the perturbative mode within the EFT with respect to that of a free scalar mode. Thus we define the phase shift such that $\lim\limits_{M\rightarrow\infty}\delta=0$ as 
\begin{equation}
	\delta=-\frac{k}{2H} \int_{-\infty}^{T_0}\left(\sqrt{W_k}- \sqrt{W^\text{dS}_k} \right)\mathrm{d} T \ , \label{eq:phaseshift}
\end{equation}
where the function $W$ for the free scalar is 
 \begin{equation}
 W^\text{dS}_k(T)=\left(1 -\frac{2H^2}{k^2T^2}  \right)\ .
  \end{equation}
With this, we find that the dimensionless spatial shift is given by
\begin{align}
\frac{k}{a(T_0)}&(a(T_0)\Delta r)= \nonumber \\
\frac{k}{H} \partial_k&\left(k \int_{-\infty}^{T_0}\left(\sqrt{W_k}- \sqrt{W^\text{dS}_k} \right)\mathrm{d} T\right)\ ,
\end{align}
where we have considered physical distances and wavelengths that redshift with the scale factor. The generalization to FRLW spacetimes is straightforward. If a particle horizon exists, the causality requirement in Eq.~\eqref{eq:causal} tells us that the EFT modes should not travel outside of the particle horizon determined by minimally coupled photons by a resolvable amount. 

\section{Shift symmetric scalar} \label{sec:scalar}
We will now use the requirement of causal propagation as encoded in Eq.~\eqref{eq:causal} to find bounds on the Wilson coefficients of an EFT by computing spatial shifts. First, we will analyze the case of a covariant, shift symmetric scalar EFT. We will only consider operators that give non-trivial scattering in trivial backgrounds. Hence, we can ignore all contributions from cubic operators since the corresponding scattering amplitudes have to be a constant (from a potential term) or vanish. We consider the shift symmetric scalar whose Lagrangian is given by
\begin{align}
	\mathcal{L}&=-\frac{1}{2}(\phi_{; \mu})^2+\frac{g_8}{M^4}((\phi_{; \mu})^2)^2 -g_{\text {matter }} \phi J(t) \nonumber \\
 & +\frac{g_{10}}{M^6}(\phi_{; \mu})^2\left[\left(\phi_{; \mu \nu}\right)^2-(\nabla^2 \phi)^2\right] +\frac{g_{12}}{M^8}\left(\left(\phi_{; \mu \nu}\right)^2\right)^2\ , \label{eq:Lag}
\end{align}
where the external source $J(T)$ sources a time-dependent background, localized around $T\simeq-\mathcal{O}(1)$, of the form
\begin{equation}
	\bar{\phi}(T)=\bar{\Phi}_0\left(\sum_{i=2}^{n}\frac{a_i}{T^i}\right) E^{-\frac{1}{ |T|^p}} \ , \label{eq:background}
\end{equation}
with $p \in \mathrm{Z}$ and $a_i=\mathcal{O}(1)$. Here, $(\phi_{; \mu})^2=\nabla_\mu\phi\nabla^\mu\phi$ and similarly for higher derivative orders. We have included the $(\phi_{; \mu})^2(\nabla^2 \phi)^2$ term which gives trivial scattering in flat space since its contribution to the spatial shift vanishes and its presence will allow us to make contact with curved spacetime generalizations of Galileons. As required for a well-defined phase shift at $T_0\lesssim-\sqrt{2} H/k$, we have $\phi(T)\rightarrow0$ as we approach the $T_0$ and $T=-\infty$. We consider a mode $\varphi=\phi-\bar{\phi}$ moving around this background and use the WKB approximation to compute the phase shift it feels at the time slice $T=T_0$ with respect to that of a free scalar in de Sitter. To do so, we construct a second-order equation of motion for the mode perturbatively. That is, we remove higher derivative contributions by using the solution to the equations of motion at lower order. Overall, we have to make sure that our perturbative approach, the WKB approximation, and the EFT expansion are under control. This can be measured by considering the following small parameters\footnote{To be more precise, these parameters are time dependent. The validity of the EFT requirement is in fact $|\epsilon_1 T \phi'(T)|\ll 1$, but since $\phi$ is localized near $T=-1$, requiring $\epsilon_1\ll 1$ is enough. Meanwhile, the WKB approximation requires $\epsilon_3\ll |T|<|T_0|\ll 1$, where this choice of $T_0$ allows us to probe deeper into the de Sitter bulk.}
\begin{equation}
\epsilon_1 \equiv \frac{H \bar{\Phi}_0}{ M^2} \ll 1, \quad \epsilon_2 \equiv \frac{k H}{M^2} \ll 1, \ \text { and } \
\epsilon_3 \equiv \frac{H }{k} \ll 1 \ , \label{eq:eps}
\end{equation}
which encode the WKB approximation ($\epsilon_3\ll1$) and the validity of the EFT ($\nabla\phi/M\sim \epsilon_1 \ll1 \ , \  \nabla/M\sim \sqrt{\epsilon_2\epsilon_3}\ll 1 $). The contributions of each Wilson coefficient are of order:
$g_8: \mathcal{O}\left(\epsilon_1^2\right)$, $g_{10}: \mathcal{O}\left(\epsilon_1^2 \epsilon_2 \epsilon_3\right)$, $g_{12}: \mathcal{O}\left(\epsilon_1^2 \epsilon_2^2 \right)$, so that the perturbative equation of motion that we use is valid as long as 
\begin{equation}
	g_8<\left(\epsilon_1^2\right)^{-1}, \quad g_{10}< \left(\epsilon_1^2 \epsilon_2 \epsilon_3\right)^{-1}, \quad g_{12}< \mathcal{O}\left(\epsilon_1^2 \epsilon_2^2 \right)^{-1} \ .
\end{equation}
Higher-order contributions in $\epsilon_1$ encode operators with more fields and higher-order contributions in $\epsilon_2, \  \epsilon_3$ arise from higher order WKB corrections as well as operators with more derivatives.

We can now compute the phase shift experienced by this perturbation with respect to a free particle in de Sitter which is given in Eq.~\eqref{eq:phaseshift} with
\begin{align}
W_k=&1-\frac{2 \epsilon _3^2}{T^2}-4 g_{10} T^2 \epsilon _1^2 \epsilon _2 \epsilon _3 \bar{\phi} '
\left(6T \bar{\phi} ''+7 \bar{\phi} '\right) \nonumber \\
&-8 g_8 T^2
\epsilon _1^2 \bar{\phi}'^2-8 g_{12} T^4 \epsilon _1^2 \epsilon _2^2 \left(T \bar{\phi} ''+2 \bar{\phi} '\right)^2 \nonumber\\
&-12 g_8 \epsilon _1^2 \epsilon _3^2 \left(T^2 \bar{\phi}
''^2-\bar{\phi}'^2+T \bar{\phi} ' \left(T \bar{\phi} ^{(3)}+2 \bar{\phi} ''\right)\right) \ . \label{eq:scalarW}
\end{align}
Here we have considered all terms of order $\epsilon_1^2\epsilon_{2/3}^2$ and neglected higher-order contributions such as $\epsilon_1^4$ that encode operator with more than $4$ fields and also terms with an additional $\epsilon_{2/3}$ suppression. This implies that our power counting requires $\epsilon_1<\epsilon_2,\epsilon_3$. We will refer to this as the leading order contribution, that is, the terms of the following order:
\begin{equation*}
    \text{LO}: \quad \{\epsilon_1^2\ , \epsilon_1^2 \epsilon_2^2\ , \epsilon_1^2 \epsilon_3^2 \ , \epsilon _1^2 \epsilon_2\epsilon_3\}\sim \mathcal{O}\left(10^{-2}-10^{-1}\right) \ ,
\end{equation*}
and the next-to-leading order terms are
\begin{align*}
    \text{NLO}: \quad \{&\epsilon_1^4\ , \epsilon_1^4 \epsilon_2^2\ , \epsilon_1^4 \epsilon_3^2 \ , \epsilon _1^2 \epsilon_4\epsilon_3 \ , \nonumber\\
    &\epsilon_1^2 \epsilon_2^2\epsilon_2^2\ , \epsilon_1^2 \epsilon_2\epsilon_2^3 \ , \epsilon_1^2 \epsilon_2^3\epsilon_2\}\sim \mathcal{O}\left(10^{-3}\right) \ .
\end{align*}
The spatial shift including the next-to-leading order terms can be found in Eq.~\eqref{eq:DrNLO}.  While there are non-sign definite $g_8$ terms, these have to be suppressed to have well-defined EFT and WKB expansions. Then, having $g_8>0$ ($g_8<0$) gives a negative (positive) spatial shift. Meanwhile, the $g_{10}$ term does not seem to be sign definite, but its contribution to the spatial shift is a positive definite term plus a total derivative. For the assumed one-dimensional, time-dependent background, the contributions to the spatial shift from all the $g_i$ considered here have definite sign, so we can only bound the Wilson coefficients from one side. 

We obtain bounds on a given coefficient by optimizing the background solution to give a larger than naively expected contribution to the spatial shift from its corresponding operator. This is done by choosing $T_0$, $a_i$, and $p$ in Eq.~\eqref{eq:background} such that some derivatives grow while the other ones stay of order one. To confirm that the EFT and WKB expansions are under control, in addition to imposing the bounds in Eq.~\eqref{eq:eps}, we explicitly compute the corrections from the NLO operators and verify that they are suppressed for our choice of background, that is, for the choice of $T_0$, $a_i$ and $p$. Details on this computation can be found in Appendix \ref{ap:ho}. We find that this optimization leads to the choice $n=4$ and $p=4$ for our expansion in Eq.~\eqref{eq:background} so that higher order derivatives do not grow uncontrollably at small $T$ and at the same time, $\bar{\phi}(T)\rightarrow0$ and  $\bar{\phi}^{(n)}(T)\rightarrow0$ at the time slice $T=T_0$. 

\begin{figure}
	\includegraphics[scale=0.3]{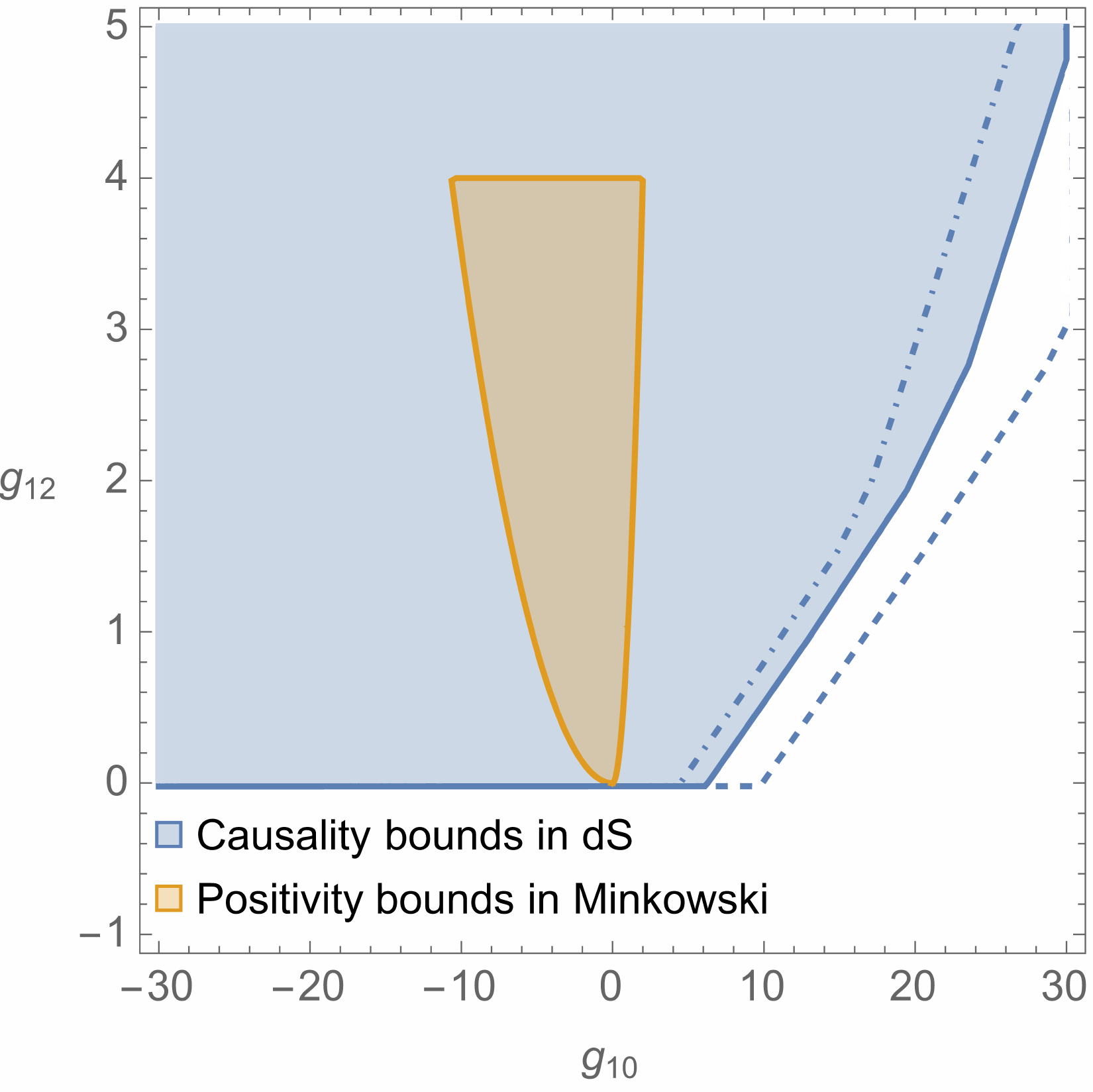}
	\caption{Comparison of causality bounds on scalar shift-symmetric theories from propagation in de Sitter (blue) against positivity bounds obtained from scattering in flat space (yellow) \cite{Tolley:2020gtv}. The colored regions are the allowed ones by the corresponding criteria. The solid line corresponds to the criteria in Eq.~\eqref{eq:causal}, the dashed-dotted line to the same criteria but with $0.5$ on the RHS, and the dashed one with a value of $2$.}
	\label{fig:boundsShiftScalar}
\end{figure}

As in the flat space case, the simplest bound is $g_8>0$. This can be inferred easily by requiring a power counting where $\epsilon_2,\epsilon_3\ll\epsilon_1$ such that
\begin{align*}
    \text{LO}:& \quad \{\epsilon_1^2\}\sim \mathcal{O}\left(10^{-1}\right) \ ,\\
    \text{NLO}:& \quad \{\epsilon_1^2 \epsilon_2^2\ , \epsilon_1^2 \epsilon_3^2 \ , \epsilon _1^2 \epsilon_2\epsilon_3\}\sim \mathcal{O}\left(10^{-2}-10^{-3}\right) \ .
\end{align*}
In this case, only the $\epsilon_1^2$ term is considered leading order while all other terms are subleading. We have checked explicitly that we can construct backgrounds leading to this result\footnote{The strict positivity of $g_8$ is obtained since numerically the specific lower bound is smaller than the numerical precision considered for the calculation. Similarly, the contributions of higher-order operators fall in this category.}. Thus, we can set $g_8=1$ which simply rescales the precise relation between the EFT cutoff and the scale $M$. The bounds that we obtain on $g_{10}$ and $g_{12}$ are shown in Fig.~\ref{fig:boundsShiftScalar}. The bound on $g_{12}$ is
\begin{equation}
    g_{12}>0 \ , \label{eq:g12}
\end{equation}
but it becomes stronger as $g_{10}$ is increased. The bound on Eq.~\eqref{eq:g12} holds for order one numbers on the RHS of Eq.\eqref{eq:causal}. We do not quote a similar bound on $g_{10}$ since it is of order $100$ and hence not significant within the regime of validity of the EFT.

Besides the operators in Eq. \eqref{eq:Lag}, we can also include a term with $6$ fields at mass dimension $12$ which is given by
\begin{align}
	\mathcal{L}^\text{HO}&=\frac{h_{12}}{M^8}\left(\left(\phi_{; \mu }\right)^2\right)^3\ . \label{eq:LagHO}
\end{align}
To probe this operator, we will include corrections to the spatial shift of order $\epsilon_{i}^4$, where $i=1,2,3$, and neglected higher-order contributions with additional $\epsilon_i$ suppression. The power counting is now
\begin{align*}
    \text{LO}: \quad \{&\epsilon_1^2\ , \epsilon_1^2 \epsilon_2^2\ , \epsilon_1^2 \epsilon_3^2 \ , \epsilon _1^2 \epsilon_2\epsilon_3\ , \epsilon_1^4\}\sim \mathcal{O}\left(10^{-2}-10^{-1}\right), \\
    \text{NLO}: \quad  \{&\epsilon_1^4\ , \epsilon_1^4 \epsilon_2^2\ , \epsilon_1^4 \epsilon_3^2 \ , \epsilon _1^2 \epsilon_4\epsilon_3 \ , \nonumber\\
    &\epsilon_1^6\ , \epsilon_1^2 \epsilon_2^2\epsilon_2^2\ , \epsilon_1^2 \epsilon_2\epsilon_2^3 \ , \epsilon_1^2 \epsilon_2^3\epsilon_2\}\sim \mathcal{O}\left(10^{-3}\right) \ .
\end{align*}
The new contributions to $W_k$ at order $\epsilon_1^4$ are:
\begin{equation}
  W_k^{\epsilon _1^4} = \epsilon _1^4 \left(96 g_8^2 +24 h_{12} \right)T^4 \bar{\phi} '(T)^4\ .
\end{equation}
Since the contribution from $h_{12}$ to the spatial shift is positive, we can only impose an upper bound. This upper bound is not highly sensitive to changes in $g_{10},g_{12}$, which in this case, we can always suppress with the choice of a specific background. The bound on $h_{12}$ is 
\begin{equation}
 h_{12}<1.64 \ , \label{eq:h12}
\end{equation}
and becomes slightly stronger as $g_{10},g_{12}\rightarrow0$, to be more specific,  at $g_{10}=g_{12}=0$, we find $ h_{12}<1.54$. If we vary the order one number on the RHS of Eq.~\eqref{eq:causal} we find that the bound changes as follows: when the number is changed to $0.5$ the bound is $ h_{12}<1.60 $ and when it is changed to $2$ it is $ h_{12}<1.72 $. Note that bounds on operators with $6$ fields cannot be obtained from the standard positivity bounds using tree-level $2-2$ scattering. An extension to higher-point positivity bounds for $P(X)$ theories has been considered in \cite{Chandrasekaran:2018qmx}. Interestingly, within this context they find that the operator $h_{12}$ is required to be negative.

It is worth highlighting that, although not obvious in Eq.~\eqref{eq:scalarW}, the $g_{10}$ term is proportional to the Hubble parameter, so that it vanishes in the flat space limit, as analyzed in \cite{CarrilloGonzalez:2022fwg}. While this simple time-dependent background would give no bounds on $g_{10}$ when considering propagation around a Minkowski spacetime, it gives an upper bound when propagating in a de Sitter background. This is the case since in Minkowski the $g_{10}$ operator realizes the Galileon symmetry, but in curved spacetimes, this is no longer true. In our setup, if instead of measuring the phase shift at a time slice near (Hubble) horizon crossing, we measured it at a time slice with large negative $T$, we will be probing the Minkowski limit. In that case, we will obtain no bounds on $g_{10}$. Instead, one could be interested in probing generalizations of Galileons to curved spacetime \cite{Goon:2011qf,Goon:2011xf,Burrage:2011bt,Deffayet:2009wt}. Here, we will briefly mention the bounds on one of them and in the next section, we analyze the second case in more detail.

\paragraph{Quartic Covariant Galileon}
The covariant Galileon \cite{Deffayet:2009wt} removes higher derivatives in all field equations at the cost of losing the Galileon symmetry. Nevertheless, the non-minimal couplings to gravity only lead to a weakly broken galileon symmetry \cite{Pirtskhalava:2015nla}. This type of theories correspond to a subset of the broader Horndeski class \cite{Horndeski:1974wa}. On a fixed de Sitter background, the covariant Galileon corresponds to taking $g_8=3 g_{10} \epsilon_2 \epsilon_3$. Thus, the leading term is now $g_{10}$, and it gives a negative contribution to the spatial shift 
\begin{equation}
    k\Delta r= -8 g_{10} \epsilon_1^2 \epsilon_2 \int_{-\infty}^{T_0}\ud T \  T^2 \bar{\phi}'(T)^2 \ ,
\end{equation}
By choosing $\epsilon_3\ll\epsilon_2,\epsilon_1$, we can enhance the $g_{10}$ contribution and it is easy to find backgrounds $\bar{\phi}(T)$ leading to the bound
\begin{equation}
	g_{10}>0 \ ,
\end{equation}
where $g_{10}$ is now the Wilson coefficient in front of the quartic covariant Galileon.

\section{de Sitter Galileon} \label{sec:dSgal}
The de Sitter Galileon \cite{Goon:2011qf,Burrage:2011bt} realizes the symmetry breaking pattern $\mathfrak{i s o}(D, 1) \rightarrow \mathfrak{s o}(D, 1)$ in the non-relativistic limit, that is, is the generalization of the flat space Galileons to the de Sitter case. The Lagrangian changes by a total derivative under the shift
\begin{equation}
    \pi \longrightarrow \pi-\frac{1}{H \tau}\left(c+v_i x^i+ v_0 H \left(x^i x_i-\tau^2\right)\right)\ , \label{eq:dSgalSym}
\end{equation}
where $v^i$ is a constant 3-vector, and $v_0$ and $c$ are constants. For the kinetic term to be invariant (up to total derivatives) under this shift, one has to consider a massive scalar with mass $ m_\pi^2=-4H^2$. Notice that the coupling of the background to an external source in Eq.~\eqref{eq:Lag} breaks this symmetry, but as is standard, one can consider a gravitational coupling which is Planck mass suppressed so that this is only softly broken. More details on this type of source can be found in Appendix \ref{ap:source}.

In this case, we cannot neglect the cubic terms since the theory cannot be defined around flat space. The EFT power counting for the de Sitter Galileon requires 
\begin{equation}
\tilde{\epsilon}=\epsilon_1\sqrt{\epsilon_2\epsilon_3}\ll1 \ , \quad  \epsilon_3 \ll 1
\end{equation}
where these expansion parameters have been defined in Eq.~\eqref{eq:eps} and the first equation encodes the EFT expansion while the second one corresponds to the WKB expansion. The $n-$th de Sitter Galileon contributes at order $\tilde{\epsilon}^{n-2}\epsilon_3^m$, where $m\geq0$. Following the same procedure as above, we bound the Wilson coefficients of the cubic and quartic Galileons by considering contributions to the spatial shift up to order $\tilde{\epsilon}^2$ and neglecting WKB corrections of order $\tilde{\epsilon}\epsilon_3^2$ and higher. This means that we take $\epsilon_3^2\ll \tilde{\epsilon}\ll 1$. The order of magnitude of the contributions at leading and next-to-leading order are thus
\begin{align*}
    \text{LO}:& \quad \{\tilde{\epsilon}\ ,\tilde{\epsilon}^2 \ , \tilde{\epsilon}\epsilon_3\}\sim \mathcal{O}\left(10^{-1}-10^{-2}\right) \ ,\\
    \text{NLO}:& \quad \{\tilde{\epsilon}^3 \ ,\tilde{\epsilon}^2\epsilon_3 \ , \tilde{\epsilon}\epsilon_3^2 \}\sim \mathcal{O}\left(10^{-3}\right) \ .
\end{align*}
At leading order the terms in the Lagrangian contributing to the spatial shift are
\begin{align}
    \lag^\text{dS Gal}&=\lag^\text{dS Gal}_2+\lag^\text{dS Gal}_3+\lag^\text{dS Gal}_4 \ ,\\
    \lag^\text{dS Gal}_2&=-\frac{1}{2}\left((\pi_{;\mu})^2-4H^2\pi\right) \ ,\\
    \lag^\text{dS Gal}_3&= \frac{f_8}{M^3}\left((\pi_{;\mu})^2 \nabla^2 \pi+6 H^2 \pi(\pi_{;\mu})^2-8 H^4 \pi^3\right) \ ,\\
    \lag^\text{dS Gal}_4&= \frac{g_{10}}{M^6} \left((\pi_{;\mu})^2\left[\left(\pi_{; \mu \nu}\right)^2-(\nabla^2 \pi)^2\right]-\frac{1}{2} H^2(\pi_{;\mu})^4\right. \nonumber\\
    &\left.-6 H^2 \pi(\pi_{;\mu})^2 \nabla^2 \pi -18 H^4 \pi^2(\pi_{;\mu})^2+12 H^6 \pi^4 \right) \ .
\end{align}
The explicit expressions for the equations of motion and spatial shift can be found in Appendix \ref{ap:dSGal}. 

\begin{figure}
	\includegraphics[scale=0.3]{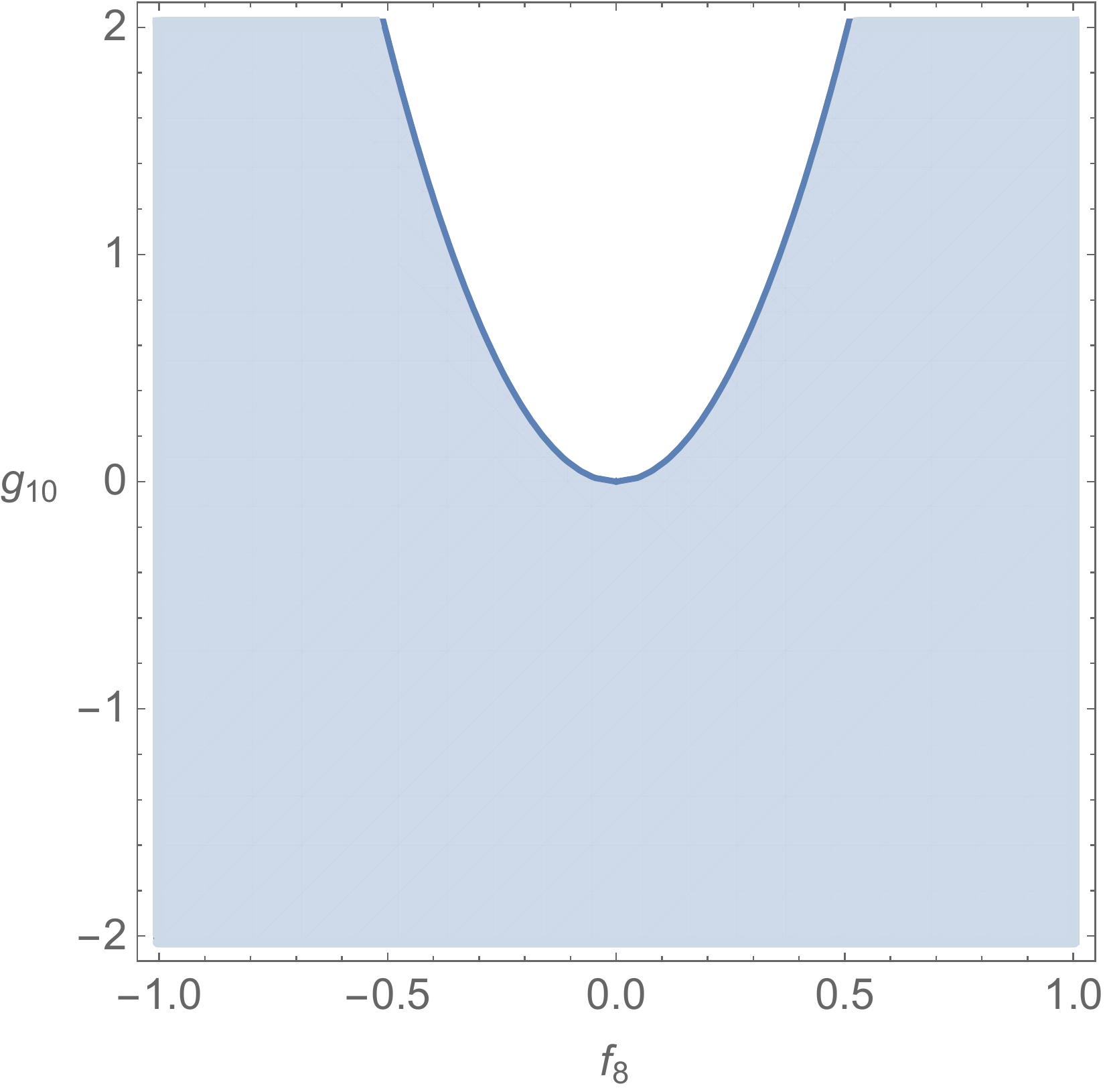}
	\caption{Causality bounds on the de Sitter Galileon at cubic ($f_8$) and quartic ($g_{10}$) order. The blue region is allowed by the requirement in Eq.~\eqref{eq:causal}. The bound does not change if we change the $\mathcal{O}\left(1\right)$ number on the right-hand side of Eq.~\eqref{eq:causal}.}
	\label{fig:boundsdSGal}
\end{figure}

The $f_8$ term will have no contribution at linear order since at that order it gives a total derivative in the integrand of the spatial shift and the other linear contribution is suppressed by the WKB expansion parameter. Meanwhile, the linear in $g_{10}$ the terms with $H^4$ and $H^6$ factors do not contribute since they contribute to the effective potential and thus are inherently suppressed. The usual quartic Galileon and the $H^2$ term contribute to the spatial shift with a positive-definite contribution and a total derivative so that
\begin{equation}
    k\Delta r^{g_{10}}= 6 g_{10} \epsilon_1^2 \epsilon_2 \int_{-\infty}^{T_0}\ud T \ T^2 \bar{\phi}'(T)^2 \ ,
\end{equation}
which leads to an upper bound. Optimizing the background to find the strongest bounds we get 
\begin{equation}
g_{10}<7.81 f_8^2
\end{equation}
which can be observed in Fig.~\ref{fig:boundsdSGal}. Note that at $f_8=0$ we have the opposite sign to that of the covariant Galileon.

\section{A Note on Potentials} \label{sec:pot}
Scalar fields with potentials are relevant for various cosmological scenarios from early to late Universe. One could ask whether potential terms can lead to acausal propagation since they can modify the 2-point correlator around non-trivial backgrounds. Let us consider a canonically normalized scalar field with a potential $V(\phi)$ living in a fixed FRLW background. The perturbations around a time-dependent scalar background $\bar{\phi(t)}$ generated by a localized external source will acquire a time-dependent effective mass $m_\text{eff.}(\bar{\phi(t)})$. The phase shift is thus given by
\begin{equation}
	\delta=-\frac{k}{2H} \int_{-\infty}^{T_0}\left(\frac{m^2_\text{eff.}(\bar{\phi}(T))}{2 k^2} \right)\mathrm{d} T \ ,\label{eq:phaseshiftPot}
\end{equation}
where $H=H(T_0)$ is the Hubble parameter at $T_0$. We can estimate the spatial shift to be of order
\begin{equation}
     k\Delta r\simeq-\frac{m^2_\text{eff.}(\Phi_0)}{k^2}\frac{1}{\epsilon_3} \ ,
\end{equation}
which automatically satisfies the causality requirement as long as we have a positive effective mass squared. In the case $m^2_\text{eff.}<0$, it will naively seem that we can get acausal propagation, but we should be more careful. The scale $m^2_\text{eff.}$ is set by the background equation of motion of the scalar which has an external localized source. Thus, we have two options. First, if the derivatives dominate we must have $|m^2_\text{eff.}|\lesssim H^2$ which implies an unresolvable spatial advance $ |k\Delta r|\lesssim \epsilon_3<1$. If instead, the potential dominates, then $|m^2_\text{eff.}|> H^2$ and the scale of variation of the background is not set by $H$, but rather by $|m^2_\text{eff.}|^{1/2}$. The validity of the WKB approximation implies again an unresolvable spatial advance: $ |k\Delta r|\lesssim |m^2_\text{eff.}|/k<1$. Hence, the potential terms can not lead to acausal propagation. Note that outside of the WKB regime of validity, it could be possible that potential terms lead to a resolvable spatial advance, but this is not a violation of causality, but rather an effect from bouncing off the potential barrier similar to that analyzed in non-relativistic systems \cite{Eisenbud,Wigner:1955zz}; for a review on this topic see \cite{Mizera:2023tfe} and a field theory example in Appendix A of \cite{CarrilloGonzalez:2022fwg}.

\section{Discussion} \label{sec:disc}
We have shown that the requirement of causal propagation imposes bounds on the Wilson coefficients of higher-order operators for EFTs in de Sitter. Our causality criterion consists of requiring that the EFT modes do not propagate further than a minimally coupled photon mode by a resolvable amount. To bound higher order terms which naively are always subleading, we constructed non-trivial scalar field backgrounds that enhance a specific operator while keeping the higher order terms suppressed. 

Our point of view is that a theory should have causal propagation around any localized background that can be continuously deformed to the trivial one ($\phi=0$). We have been agnostic about the precise origin of the external source and have only required it to be localized. If the scalar field was describing a component of a complex scalar coupled to a $U(1)$ gauge field then any positive or negative source would be physical since it would simply encode the sign of the charge. On the other hand, if the field had gravitational couplings, one should be more careful and make sure that the type of matter sourcing the profile is not pathological. In Appendix \ref{ap:source}, we analyzed this case and showed that our profile can be sourced by a stress-energy tensor satisfying the weak energy condition.

We also want to highlight that the bounds obtained here are certainly not optimal. Likely, a different profile and a better optimization procedure for maximizing the spatial shift can lead to stronger bounds. Similarly, considering backgrounds that break additional symmetries can lead to stronger bounds. Here we have only considered backgrounds consistent with the FRLW symmetries.

In the analysis above, we have kept the de Sitter background fixed. It would be interesting to understand how dynamical gravity, that is, both the interaction with propagating graviton modes and the backreaction on the background, can affect any of these calculations. It is worth noting that the expected corrections will be Planck mass suppressed. We have worked with fully covariant theories with symmetry breaking due to their time-dependent background. In a similar direction, one could be interested in exploring the bounds between coefficients of other exceptional field theories in de Sitter space such as those in \cite{Hinterbichler:2022vcc,Bonifacio:2021mrf}.  Another interesting direction is to apply this technique to bound higher dimensional gravitational operators in the spirit of the leading order analysis in \cite{deRham:2020zyh}.  More generally, we would like to impose bounds on an EFT defined around FRLW backgrounds such as the EFT of inflation. This would be the subject of future work.

\section*{Acknowledgements}
MCG would like to thank Paolo Benincasa, Claudia de Rham , Andrew Tolley, and Sebasti\'an C\'espedes as well as the organizers and participants of the Workshop on Scattering Amplitudes and Cosmology at the ICTP for insightful discussions. During the completion of this work, the research of MCG  was supported by the Imperial College Research Fellowship and the STFC grant ST/T000791/1.

\onecolumngrid
\appendix

\section{Higher-order correction to shift-symmetric scalar}  \label{ap:ho}
In this appendix, we show the higher-order contributions to the phase shift and hence the spatial shift that we analyze to make sure that both the WKB and EFT expansions are under control. At next-to-leading order (NLO), that is, including the contributions at order $\epsilon_1^4$ and $\epsilon_1^2 \epsilon_{2/3}^4$, we have new operators contributing to the spatial shift, their Lagrangian reads
\begin{equation}
	\lag\subset \frac{h_{12}}{M^8}(\nabla \phi)^6+\frac{h_{14}}{M^{10}}(\nabla \phi)^4(\phi;_{\mu \nu})^2+\frac{g_{14}}{M^{10}}(\phi;_{\mu \nu})^2(\phi;_ {\alpha \beta \gamma})^2 \  .
\end{equation}
Their contribution to the equations of motion is encoded in the function $W$ appearing in Eq.~\eqref{eq:EOM} which, including these new operators and all the contributions at NLO, now reads 
{ \small
\begin{align}
W&=1-\frac{2 \epsilon _3^2}{T^2}+g_8 \left(\epsilon _1^2 \left(-8 T^2 \bar{\phi}'^2-12 \epsilon _3^2 \left(T^2 \bar{\phi} ''^2-\bar{\phi}'^2+T \bar{\phi} ' \left(T \bar{\phi}
   ^{(3)}+2 \bar{\phi} ''\right)\right)\right)
   +96 g_{10} T^4 \epsilon _1^4 \epsilon _2 \epsilon _3 \bar{\phi} '^3 \left(3T \bar{\phi} ''+5\bar{\phi} '\right) \right. \nonumber \\
   &  +   32 g_{12} T^6 \epsilon _1^4 \epsilon _2^2 \bar{\phi}'^2\left(T \bar{\phi} ''+2 \bar{\phi} '\right) \left(7 T \bar{\phi}
   ''+10 \bar{\phi} '\right)-528 h_{12} T^6 \epsilon _1^6 \bar{\phi}'^6 \Big)+g_8^2 \epsilon _1^4 \left(96 T^4 \bar{\phi} '^4+144 T^3 \epsilon _3^2 \bar{\phi}'^2\left(2 T \bar{\phi} ''^2+\bar{\phi} '
   \left(T \bar{\phi} ^{(3)}+4 \bar{\phi}''\right)\right)\right) \nonumber \\
   &+h_{12} \epsilon _1^4 \left(24 T^4 \bar{\phi} '^4+60 T^2 \epsilon _3^2 \bar{\phi}'^2\left(3 T^2 \bar{\phi} ''^2+ \bar{\phi}
   '^2+T \bar{\phi} ' \left(T \bar{\phi} ^{(3)}+6 \bar{\phi} ''\right)\right)\right)+ g_{10} \epsilon _1^2 \epsilon _2 \epsilon _3\left(-4 T^2  \bar{\phi} ' \left(6T \bar{\phi} ''+7 \bar{\phi} '\right) \right. \nonumber \\
   &\left. -18
   \epsilon _3^2 \left(T^2 \bar{\phi} ''^2- \bar{\phi}'^2+T \bar{\phi} ' \left(T\bar{\phi}^{(3)}+2\bar{\phi}''\right)\right)\right)+g_{12} \epsilon _1^2 \epsilon _2^2 \left(-8T^4 \left(T \bar{\phi}''+2 \bar{\phi} '\right)^2  +8 T^3 \epsilon _3^2 \left( \bar{\phi} ' \left(T \left(T \bar{\phi} ^{(4)}+18 \bar{\phi} ^{(3)}\right)+48 \bar{\phi} ''\right)\right.\right. \nonumber \\
   &
  \left.  \left. + T \left(T^2 \left(\bar{\phi} ^{(3)}\right)^2+T \left(T \bar{\phi} ^{(4)}+19 \bar{\phi} ^{(3)}\right) \bar{\phi} ''+40 \left(\bar{\phi}
   ''\right)^2\right)\right)\right)-1152 g_8^3 T^6 \epsilon _1^6 \left(\bar{\phi} '\right)^6 \nonumber \\
   &-4 g_{14} T^4 \epsilon _1^2 \epsilon _2^3 \epsilon _3 \left(T^2 \left(T^2 \left(\bar{\phi} ^{(3)}\right)^2+T \left(T \bar{\phi} ^{(4)}+6 \bar{\phi} ^{(3)}\right) \bar{\phi} ''-28 \left(\bar{\phi}
   ''\right)^2\right)+4 T \bar{\phi} ' \left(T \left(T \bar{\phi} ^{(4)}+6 \bar{\phi} ^{(3)}\right)-13 \bar{\phi} ''\right)-52 \left(\bar{\phi}
   '\right)^2\right) \nonumber\\
   &+ 8 h_{14} T^4 \epsilon _1^4 \epsilon _2 \epsilon _3 \left(\bar{\phi} '\right)^2 \left(5 T^2 \left(\bar{\phi} ''\right)^2+2 T
   \bar{\phi} ' \left(T \bar{\phi} ^{(3)}+5 \bar{\phi} ''\right)-7 \left(\bar{\phi} '\right)^2\right) \ .
\end{align}
}
Additionally, we have contributions arising from higher-order WKB corrections to the phase shift which are given by
\begin{equation}
\begin{aligned}
	&\delta_{\mathrm{WKB}}^{(0)}=\sqrt{\hat{W}_{\ell}},\\
	&\delta_{\mathrm{WKB}}^{(2)}=-\frac{1}{\left(\omega r_0\right)^2} \frac{1}{8 \sqrt{\hat{W}_{\ell}}}\left(\frac{\hat{W}_{\ell}^{\prime \prime}}{\hat{W}_{\ell}}-\frac{5}{4}\left(\frac{\hat{W}_{\ell}^{\prime}}{\hat{W}_{\ell}}\right)^2\right) \text {, }\\
	&\delta_{\mathrm{WKB}}^{(4)}=\frac{1}{\left(\omega r_0\right)^4} \frac{1}{32 \hat{W}_{\ell}^{3 / 2}}\left[\frac{\hat{W}_{\ell}^{(4)}}{\hat{W}_{\ell}}-7 \frac{\hat{W}_{\ell}^{\prime} \hat{W}_{\ell}^{(3)}}{\hat{W}_{\ell}^2}-\frac{19}{4}\left(\frac{\hat{W}_{\ell}^{\prime \prime}}{\hat{W}_{\ell}}\right)^2+\frac{221}{8} \frac{\hat{W}_{\ell}^{\prime \prime} \hat{W}_{\ell}^{\prime 2}}{\hat{W}_{\ell}^3}-\frac{1105}{64}\left(\frac{\hat{W}_{\ell}^{\prime}}{\hat{W}_{\ell}}\right)^4\right] \ ,
\end{aligned}
\end{equation}
where $\delta_{\mathrm{WKB}}^{(2)}$ will have contributions from $g_8$, $g_{10}$, and $g_{12}$, while $\delta_{\mathrm{WKB}}^{(4)}$ only has contributions from $g_8$ at NLO. The phase shift is now defined as
\begin{equation}
	\delta=\frac{k}{2H} \int_{-\infty}^{T_0}\left(\sum_{j \geq 0} \delta_{\mathrm{WKB}}^{(j)}-\sum_{j \geq 0} \delta_{\mathrm{WKB-dS}}^{(j)} \right)\mathrm{d} T \ . \label{eq:phaseshiftHO}
\end{equation}
from this, we can compute the spatial shift using Eq.~\eqref{eq:Dr}. Explicitly, we have
\begin{align}
   & k\Delta r=\frac{1}{\epsilon_3} \int_{-\infty}^{T_0}\ud T \left[g_8 \epsilon _1^2 \left(-4 T^2
   \left(\bar{\phi }'\right)^2+4 \epsilon _3^2 \left(T^2 \left(\bar{\phi }''\right)^2+T \bar{\phi
   }' \left(T \bar{\phi }^{(3)}+\bar{\phi }''\right)-\left(\bar{\phi
   }'\right)^2\right)\right) \right. \nonumber \\
   &-2 g_{10} T^2 \epsilon _1^2 \epsilon _2 \epsilon _3 \bar{\phi }'
   \left(7 \bar{\phi }'+6 T \bar{\phi }''\right)-12 g_{12} T^4 \epsilon _1^2 \epsilon _2^2 \left(2 \bar{\phi }'+T
   \bar{\phi }''\right)^2 +4 T^4 \epsilon _1^4 \left(10 g_8^2+3 h_{12}\right)
   \left(\bar{\phi }'\right)^4 \nonumber \\
   &-6 g_{14} T^4 \epsilon _1^2 \epsilon _2^3 \epsilon _3 \left(T^2
   \left(T^2 (\bar{\phi }^{(3)})^2-28 (\bar{\phi }'')^2+T \left(6
   \bar{\phi }^{(3)}+T \bar{\phi }^{(4)}\right) \bar{\phi
   }''\right) +4 T \bar{\phi }' \left(T \left(6 \bar{\phi
   }^{(3)}+T \bar{\phi }^{(4)}\right)-13 \bar{\phi
   }''\right)-52 \bar{\phi }'^2\right) \nonumber\\
   &-g_{10} \epsilon _1^2 \epsilon _2 \epsilon _3^3 \left(T \bar{\phi
   }' \left(3 T^2 \bar{\phi }^{(4)}+16 \bar{\phi }''+16 T
   \bar{\phi }^{(3)}\right)+T^2 \bar{\phi }'' \left(16
   \bar{\phi }''+9 T \bar{\phi }^{(3)}\right)+2 \bar{\phi
   }'^2\right) \nonumber \\
    &-8 g_8 T^6 \epsilon _1^6 \left(52 g_8^2+27 \text{h2}\right)
   \bar{\phi }'^6 \nonumber + 96 g_8 g_{12} T^6 \epsilon _1^4 \epsilon _2^2 \bar{\phi }'^2
   \left(3 T^2 \bar{\phi }''^2+10 T \bar{\phi }' \bar{\phi
   }''+8 \bar{\phi }'^2\right)   \\
   &+2 g_{12} T^2 \epsilon _1^2 \epsilon _2^2 \epsilon _3^2 \left(T^2
   \left(3 T^2 (\bar{\phi }^{(3)})^2+117 \bar{\phi }''^2+T
   \left(56 \bar{\phi }^{(3)}+3 T \bar{\phi }^{(4)}\right)
   \bar{\phi }''\right)+4 T \bar{\phi }' \left(40 \bar{\phi
   }''+T \left(15 \bar{\phi }^{(3)}+T \bar{\phi
   }^{(4)}\right)\right)+16 \bar{\phi }'^2\right) \nonumber \\
   &-\frac{3 g_8 \epsilon _1^2 \epsilon _3^4 \left(\bar{\phi }'
   \left(T^3 \left(5 \bar{\phi }^{(4)}+T \bar{\phi
   }^{(5)}\right)-8 T \bar{\phi }''\right)+T^3 \left(\left(15
   \bar{\phi }^{(3)}+4 T \bar{\phi }^{(4)}\right) \bar{\phi
   }''+3 T (\bar{\phi }^{(3)})^2\right)+4 \bar{\phi
   }'^2\right)}{T^2} \nonumber \\
   &-2 T^2 \epsilon _1^4 \epsilon _3^2 \bar{\phi }'^2 \left(T^2
   \left(20 g_8^2+27 \text{h2}\right) \bar{\phi }''^2+T \bar{\phi
   }' \left(2 \left(8 g_8^2+21 \text{h2}\right) \bar{\phi
   }''+3 T \left(4 g_8^2+3 \text{h2}\right) \bar{\phi
   }^{(3)}\right)+\left(3 \text{h2}-16 g_8^2\right) \bar{\phi
   }'^2\right) \nonumber \\
   &+\epsilon _1^4 \epsilon _2 \epsilon _3 \left(8 g_8 g_{10} T^4
   \bar{\phi }'^3 \left(23 \bar{\phi }'+12 T \bar{\phi
   }''\right)+4 h_{14} T^4 \bar{\phi }'^2 \left(5 T^2
   \bar{\phi }''^2+2 T \bar{\phi }' \left(5 \bar{\phi }''+T
   \bar{\phi }^{(3)}\right)-7 \bar{\phi }'^2\right)\right)
   \left . \right]\ . \label{eq:DrNLO}
\end{align}
Note that the leading order $g_{10}$ term can be written as $-2\partial_T\left(3T^3\phi'^2\right)+4 T^2 \phi'^2$, where the total derivative term does not contribute leaving a positive definite contribution.

\section{de Sitter Galileon equation of motion and spatial shift}  \label{ap:dSGal}
The equation of motion for the perturbation around a time-dependent background $\bar{\phi}(t)$ is given by Eq.~\eqref{eq:EOM} with
\begin{align}
W_k=&1-\frac{6 \epsilon _3^2}{T^2}+f_8 \left(6 \epsilon _3^2 \tilde{\epsilon } \left(-\frac{16 \bar{\phi} }{T^2}+T \bar{\phi} ^{(3)}-\frac{6
   \bar{\phi} '}{T}-2 \bar{\phi} ''\right)+4 T \tilde{\epsilon } \left(T \bar{\phi} ''+2 \bar{\phi} '\right)\right) \nonumber \\
   &+48 f_8^2 T \tilde{\epsilon }^2 \left(T \bar{\phi} '+2 \bar{\phi} \right) \left(T \bar{\phi} ''+2 \bar{\phi} '\right)-24 g_{10} T \tilde{\epsilon }^2 \left(T \bar{\phi} '+\bar{\phi} \right) \left(T \bar{\phi} ''+2 \bar{\phi} '\right) \ .
\end{align}
The spatial shift thus reads
\begin{align}
k \Delta r=\frac{1}{\epsilon_3}\int_{-\infty}^{T_0} \ud T \left[f_8 \left(2 T \tilde{\epsilon } \left(T \bar{\phi} ''(T)+2
   \bar{\phi} '(T)\right)+\epsilon _3^2 \tilde{\epsilon } \left(\frac{1}{2} T^2
   \bar{\phi} ^{(4)}(T)+\frac{48 \bar{\phi} (T)}{T^2}+3 \bar{\phi} ''(T)+\frac{6
   \bar{\phi} '(T)}{T}\right)\right)\right. \nonumber \\
   \left. -12 g_{10} T \tilde{\epsilon }^2 \left(T \bar{\phi} '(T)+\bar{\phi} (T)\right)
   \left(T \bar{\phi} ''(T)+2 \bar{\phi} '(T)\right)-2 f_8^2 T \tilde{\epsilon }^2 \left(T \bar{\phi} ''(T)+2 \bar{\phi}
   '(T)\right) \left(T \left(T \bar{\phi} ''(T)-10 \bar{\phi} '(T)\right)-24
   \bar{\phi} (T)\right)\right]
\end{align}
Note that the leading order $f_8$ term can be written as $\partial_T\left(2T^2 \bar{\phi}'\right)$, that is, it is a total derivative that won't contribute to the spatial shift. 

\section{Physical requirements on the source} 
\label{ap:source}
We have considered generic external sources for the background profiles of the fields. One could ask whether this source is unphysical and hence generates acausal propagation, i.e. the problem lies in the source and not in the EFT operator. Here we show that the profiles considered in Eq. \eqref{eq:background}, can arise from a coupling to a stress-energy tensor satisfying the null energy condition. We assume that the coupling to source $J$ is given by
\begin{equation}
    g_m \bar{\phi} J= g_m \bar{\phi} \frac{T_\mu^\mu}{M_{Pl}} \ ,
\end{equation}
where $T_\mu^\mu$ is the trace of the stress-energy tensor of a perfect fluid, i.e. $T^{\mu\nu}=\text{diag.}(\rho,P,P,P)$, where $\rho=\rho(T)$ is the energy density and   $P=P(T)$ is the pressure. Conservation of the stress-energy tensor requires $P=-\rho-\rho'/(3\mathcal{H})$, where $\mathcal{H}\equiv a'/a$. We will require that this stress-energy tensor satisfies the null energy condition, that is,
\begin{equation}
 \quad \rho+P\geq0 \ .
\end{equation}
We want to see what constraints do these conditions imply on our source $T_\mu^\mu$. To do so we solve for $\rho$ in terms of $T_\mu^\mu$ by realizing that for a conserved stress-energy tensor,
\begin{equation}
 T_\mu^\mu=(3P-\rho)= -4 \rho-\frac{\rho'}{\mathcal{H}} \ .
\end{equation}
This allows us to rewrite weak energy conditions as
\begin{eqnarray}
   4 c_1 m^4 +a^4(\tau)T_\mu^\mu(\tau)+ 4 \int_1^{\tau} \ud t \ \left(a(t)^4 \mathcal{H}(t) T_\mu^\mu(t) \right)\geq 0 \ ,
\end{eqnarray}
where $c_1$ is an integration constant, $a(\tau)>0$, and $m$ is an energy scale that is fixed by choosing boundary conditions for the energy density. Given a choice of background profile, we have a choice of source and hence a given $T_\mu^\mu(t)$. We can see that as long as the constant $c_1$ is chosen such that 
\begin{equation}
c_1>-\int_1^{\tau} \ud t \ \left(a(t)^4 \mathcal{H}(t) \frac{T_\mu^\mu(t)}{m^4} \right)-a^4(\tau)\frac{T_\mu^\mu(\tau)}{4 m^4} \ ,
\end{equation}
which is always possible since the RHS is a bounded function due to the choice of a localized source, we can satisfy the null energy condition. This shows that the violations of causality observed in the analysis in the bulk are not caused by an unphysical source. 

In a similar manner, we can ask whether the stress energy tensor of the background scalar satisfies the null energy condition. This is in fact the case and it's easy to see since the stress energy tensor is dominated by the kinetic term instead of the subleading EFT corrections, thus it is approximated by the free scalar result where $\rho\sim P\sim\phi'^2/(2a^2)>0$. The EFT corrections are suppressed and do not change the sign of the energy density and pressure.

Last, we also want to verify that the backreaction of this stress-energy tensor on the metric is negligible. From the equations of motion of the background, we can estimate that $T^{\mu\nu}\sim\bar{\Phi}_0H^2M_{Pl}f^{\mu\nu}(T)$ with $f_{\mu\nu}(T)$ a diagonal matrix whose components are polynomials in T with order 1 magnitude. Thus, the backreaction on the metric is of order $\delta g_{\mu\nu}\sim (g_m \bar{\Phi}_0H^2M_{Pl}f_{\mu\nu}(T))/(M_{Pl}H)^2=g_m \bar{\Phi}_0/M_{Pl}f_{\mu\nu}(T)$. Since we can always choose $\bar{\Phi}_0\ll M_{Pl}$, we can always neglect the backreaction. Similarly, the backreaction from the background scalar on the spacetime metric can be neglected since $T^{\mu\nu}_{\bar{\phi}}\sim\bar{\Phi}_0^2H^2f^{\mu\nu}(T)$ and thus $\delta g\sim\bar{\Phi}_0^2/M_{Pl}^2\ll1$.
\bibliography{References}

\end{document}